\documentclass[showpacs,aps,prd,twocolumn,sort&compress]{revtex4}
\usepackage{amsmath}
\usepackage{epsfig}
\usepackage{subfigure}

\newcommand{\be}{\begin{equation}}\newcommand{\ee}{\end{equation}}
\newcommand{\bea}{\begin{eqnarray}}\newcommand{\eea}{\end{eqnarray}}
\newcommand{\bc}{\begin{center}}\newcommand{\ec}{\end{center}}

\def\lsim{\raise0.3ex\hbox{$\;<$\kern-0.75em\raise-1.1ex\hbox{$\sim\;$}}}
\def\gsim{\raise0.3ex\hbox{$\;>$\kern-0.75em\raise-1.1ex\hbox{$\sim\;$}}}

\begin{document}

\bigskip
\title{Effective Lagrangian for Two-photon and Two-gluon \\Decays of
  $P$-wave Heavy Quarkonium $\chi_{c0,2}$ and $\chi_{b0,2}$ states
}

\author{J.P. Lansberg$^{a}$ and
T. N. Pham$^{b}$}
\affiliation{
$^{a}$SLAC National Accelerator Laboratory, Theoretical Physics, 
Stanford University, Menlo Park, CA 94025, USA.\\
$^{b}$Centre de Physique Th\'{e}orique, CNRS \\ 
Ecole Polytechnique, 91128 Palaiseau, Cedex, France }
\date{\today}

 
\begin{abstract}
\parbox{14cm}{\rm 

In the traditional non-relativistic bound state calculation, the
two-photon  decay amplitudes of the $P$-wave
$\chi_{c0,2}$ and $\chi_{b0,2}$ states depend on the  derivative of the
wave function at the origin which can only be obtained  
from potential models. However by neglecting the relative quark momenta,
the decay amplitude can be written as the matrix element of a local
heavy quark field operator which could be obtained from other processes or
computed with  QCD sum rules technique or lattice simulation.
Following the same line as in  recent work for the 
two-photon decays of the $S$-wave $\eta_{c}$ and $\eta_{b}$ quarkonia, 
we show that the effective Lagrangian for the two-photon decays of 
the $P$-wave $\chi_{c0,2}$ and $\chi_{b0,2}$ is 
given by  the heavy quark energy-momentum tensor local operator or its
trace, the  $\bar{Q}Q$ scalar density and that  the expression for
$\chi_{c0}$  two-photon and two-gluon decay rate is 
given by the $f_{\chi_{c0}}$ decay constant and is  similar to that
 of $\eta_{c}$ which is given by $f_{\eta_{c}}$. From the existing
 QCD sum rules value for $f_{\chi_{c0}}$, we get $5\rm\,keV$ for 
the $\chi_{c0}$ two-photon width, somewhat larger than measurement,  
but possibly with large uncertainties.}

\end{abstract}
\maketitle

\section{Introduction}
With the recent new CLEO measurements\cite{CLEO,PDG} of the two-photon
decay rates of the even-parity, $P$-wave $0^{++}$ $\chi_{c0}$ and  
$2^{++}$ $\chi_{c2}$ states
and with renewed interest in radiative decays of heavy quarkonium 
states, it seems appropriate to have  another look at
the two-photon decay of heavy quarkonium from the standpoint of an
effective Lagrangian based on local operator expansion and 
heavy-quark spin symmetry, as done for the pseudo-scalar heavy quarkonia
$\eta_{c}$ and $\eta_{b}$\cite{Lansberg,Lansberg2}, for which the decay rates for
the ground state and excited states could be predicted in terms of the  
$J/\psi$ and $\Upsilon$ leptonic widths using Heavy Quark Spin Symmetry(HQSS). 
In the traditional 
non-relativisitic bound state calculation, the two-photon widths 
for the $P$-wave quarkonium state depend on the derivative
at the origin of the spatial wave function which has to be extracted from
potential models\cite{Barbieri}. Though the physics of quarkonium 
decay seems to be better understood within the conventional 
framework of QCD\cite{yr}, unlike the two-photon width of $S$-wave 
 $\eta_{c}$ and $\eta_{b}$ quarkonia which can be predicted from the 
corresponding $J/\psi$ and $\Upsilon$ leptonic widths using HQSS, there is 
no similar  prediction for the $P$-wave $\chi_{c}$ and $\chi_{b}$
states and all the existing theoretical values for the decay rates are based
on potential model calculations\cite{Barbieri,Godfrey,Barnes,Bodwin,Gupta,Munz,Huang,Ebert,Schuler,Crater,LWang,Laverty}.
To  have a prediction for the two-photon width of  
$P$-wave quarkonia, one need to express the decay amplitude in terms of
the matrix element of a heavy quark field local operator 
extracted from some known physical processes or computed in an essentially
model-independent manner, such as QCD sum rules 
technique\cite{Novikov,Colangelo} or lattice simulations\cite{Dudek}.
In fact a value of $ 438\pm 5\pm 6\rm, MeV$ for $f_{\eta_{c}}$ and 
$ 801\pm 7\pm 5\rm, MeV$ for $f_{\eta_{b}}$
consistent with the HQSS values of 
$411\rm\, MeV$ and $836\rm\, MeV$\cite{Lansberg,Lansberg2} , 
respectively, have been obtained by the lattice group TWQCD 
Collaboration\cite{Chiu} recently. With similar  determinations of  
other quarkonium decay constants, one   would be able to study  QCD 
radiative corrections and obtain 
the strong $\alpha_{s}$ coupling constant, for example, especially in 
$\chi_{b0,2}$ two-gluon decays where local operator expansion should be 
a better approximation than in $\chi_{c0,2}$ decays. In this paper,
starting from the two-photon and two-gluon 
$c\bar{c} \to \gamma\gamma,gg $ and $b\bar{b} \to \gamma\gamma,gg $
 amplitudes, we derive an  effective Lagrangian for the two-photon 
and two-gluon decays for $P$-wave quarkonium state by
neglecting the  bound state relative quark momenta compared with the
large outgoing photon or gluon momenta. We show that the decay amplitude 
is given by the heavy quark energy-momentum tensor which can be obtained
from the matrix element of its trace  as
$<0|\bar{c}\,c|\chi_{c0}>=m_{\chi_{c0}}\,f_{\chi_{c0}} $ and 
$<0|\bar{b}\,b|\chi_{b0}>=m_{\chi_{b0}}\,f_{\chi_{b0}} $. We find
that the two-photon and two-gluon decay rates of $\chi_{c0,2}$ and 
$\chi_{b0,2}$  are given in terms of
$f_{\chi_{c0}}$ and $f_{\chi_{b0}}$, similar to the $\eta_{c}$ and
$\eta_{b}$ two-photon decay rates given by $f_{\eta_{c}}$ and $f_{\eta_{b}}$.

\section{Effective Lagrangian for $\chi_{c0,2}\to \gamma\gamma$ and $\chi_{b0,2}\to \gamma\gamma$}
\begin{figure}[h]
\centering
\includegraphics[height=2.5cm,angle=0]{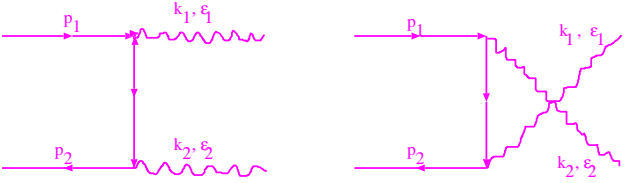}
\caption{Diagrams for $Q\bar{Q} $ annihilation to two photons.}
\label{fig1}
\end{figure}

 Following \cite{Kuhn,Guberina}, we consider the amplitude for the annihilation
of a quark and an antiquark with momentum $p_{1}$ and $p_{2}$ represented
by the diagrams  in Fig.(\ref{fig1}):
\be
{\cal A} = \bar{v}(p_{2})({\cal O}_{1}+{\cal O}_{2} )u(p_{1})
\label{amp}
\ee
with
\bea
{\cal O}_{1} &=&\frac{1}{i}\Biggl[(-ie\not\! \epsilon_{2})
i\frac{(\not\! p_{1}-\not\! k_{1} +m_{Q})}{(p_{1}-k_{1})^{2} -m_{Q}^{2}}(-ie\not\! \epsilon_{1})\Biggr] \\
{\cal O}_{2} &=& \frac{1}{i}\Biggl[(-ie\not\! \epsilon_{1})i\frac{(\not\! p_{1}-\not\! k_{2}
  +m_{Q})}{(p_{1}-k_{2})^{2} -m_{Q}^{2}}(-ie\not\! \epsilon_{2})\Biggr]
\label{O12}
\eea
where ($\epsilon_{1},k_{1}$) and ($\epsilon_{2},k_{2}$) are the 
polarizations and momenta of the outgoing photons and $m_{Q}$ the 
heavy quark mass. The total energy-momentum of the quark-antiquark 
system is the energy-momentum of the quarkonium bound state defined as 
$Q=p_{1} + p_{2}$ and  mass $M$.

Using Dirac equation and expanding ${\cal O}_{1}$ and ${\cal O}_{2}$, and
putting:
\be
 q=p_{1} - p_{2},\quad Q= k_{1} + k_{2}, \quad K=k_{1} - k_{2}
\label{pq}
\ee
we have
\bea
&&\kern -0.5cm{\cal O}_{1} =-e^{2}Q_{c,b}^{2}\frac{(\epsilon_{1}\cdot
  \epsilon_{2}(\not\!k_{1}-\not\! k_{2})-i\epsilon(\epsilon_{2},K,\epsilon_{1},\sigma)\gamma_{\sigma}\gamma_{5})/2}{[(p_{1}-k_{1})^{2}-m_{Q}^{2}]} \nonumber\\
&& \kern-0.3cm -e^{2}Q_{c,b}^{2}\frac{(-\epsilon_{2}\cdot (p_{2}+k_{1}/2)\not\!\epsilon_{1}
+\epsilon_{1}\cdot (p_{1}+k_{2}/2)\not\!\epsilon_{2})}{[(p_{1}-k_{1})^{2}-m_{Q}^{2}]}\label{O12a}\\
&&\kern -0.5cm{\cal O}_{2} =-e^{2}Q_{c,b}^{2}\frac{(\epsilon_{1}\cdot
  \epsilon_{2}(\not\!k_{2}-\not\! k_{1})+i\epsilon(\epsilon_{2},K,\epsilon_{1},\sigma)\gamma_{\sigma}\gamma_{5})/2}{[(p_{1}-k_{2})^{2}-m_{Q}^{2}]} \nonumber\\
&& \kern-0.3cm -e^{2}Q_{c,b}^{2}\frac{(\epsilon_{2}\cdot (p_{1}+k_{1}/2)\not\!\epsilon_{1}
-\epsilon_{1}\cdot (p_{2}+k_{2}/2)\not\!\epsilon_{2})}{[(p_{1}-k_{2})^{2}-m_{Q}^{2}]}
\label{O12b}
\eea
The $P$-wave $\chi_{c0,2}$ and $\chi_{b0,2}$ two-photon (two-gluon)
decay amplitudes are given by the $P$-wave part of the 
$Q\bar{Q}\to \gamma\gamma, gg $ annihilation amplitude which is given by the 
 $k_{1}\cdot q$, $\epsilon_1\cdot q$ and $\epsilon_2\cdot q$ terms
in  ${\cal O}_{1},{\cal O}_{2}$. By neglecting  term containing the
relative quark momenta $q$ in the quark propagator\cite{Kuhn} 
we find, ($Q^{2}_{c,b} $ being the heavy quark charge),
\bea
&&{\cal M}(Q\bar{Q}\to \gamma\gamma)=-e^{2}Q^{2}_{c,b}\times\nonumber \\
&&\kern-0.4cm \bar{v}(p_{2}) \biggl[k_1\cdot q(-2 \epsilon_1\cdot k_2 \not \!\epsilon_2 + 2 \epsilon_1\cdot
  \epsilon_2 \not\!k_{2} + 2 \epsilon_2\cdot k_1 \not\!\epsilon_1)+
 \nonumber\\
 &&\kern-0.4cm   M^{2}( \epsilon_2\cdot q \not\!\epsilon_1 +
 \epsilon_1\cdot q\not \!\epsilon_2 )/2\biggr]u(p_{1})\biggl[(k_{1}-k_{2})^{2}/4 -m_{Q}^{2}\biggr]^{-2}
\label{M}
\eea
which is now reduced to the matrix element of a local operator for 
two-photon or two-gluon decays of $P$-wave quarkonia with the outgoing photon 
or gluon having large momenta compared to the relative quark-antiquark 
momenta as given by the numerator of the amplitude in Eq.(\ref{M}). We 
have (rewriting $M^{2}$ as $2k_{1}\cdot k_{2}$ and 
$k_1\cdot q\epsilon_2\cdot k_1 \not\!\epsilon_1$ as
$-k_2\cdot q\epsilon_2\cdot k_1 \not\!\epsilon_1$ in Eq.(\ref{M})):
\be
{\cal M}(Q\bar{Q}\to \gamma\gamma)=-e^{2}Q^{2}_{c,b}\frac{A_{\mu\nu}\bar{v}(p_{2})T_{\mu\nu}u(p_{1})}
{[(k_{1}-k_{2})^{2}/4 -m_{Q}^{2}]^{2}}
\label{Amn}
\ee
with $A_{\mu\nu}$ the photon part of the amplitude  and the
heavy quark part $T_{\mu\nu}$ given by:
\bea
A_{\mu\nu}&=& -2\epsilon_1\cdot k_{2}\epsilon_{2\mu}k_{1\nu}+2\epsilon_1\cdot\epsilon_2 k_{2\mu}k_{1\nu}\nonumber \\
&&-2\epsilon_2\cdot k_{1}\epsilon_{1\mu}k_{2\nu}+(k_{1}\cdot k_{2})(\epsilon_{1\mu}\epsilon_{2\nu}+\epsilon_{2\mu}\epsilon_{1\nu}) \\
T_{\mu\nu}&=& (q_{1\mu}-q_{2\mu})\gamma_{\nu}
\label{T}
\eea
We see that  $\bar{v}(p_{2})T_{\mu\nu}u(p_{1}) $ is the matrix element
of $\theta_{Q\mu\nu}=\bar{Q}(\overrightarrow\partial_{\mu}-\overleftarrow\partial_{\mu})\gamma_{\nu}Q$, the heavy quark energy-momentum tensor.
The photon part can also be written in terms of the photon
field operator $F_{\mu\nu}$, but for simplicity, we will keep the 
matrix element form given by $A_{\mu\nu}$. The effective Lagrangian 
for two-photon and two-gluon
decay of $P$-wave $\chi_{c0,2}$ and $\chi_{b0,2}$ states is then given
by:
\bea
\kern-0.5cm {\cal L}_{\rm eff}(Q\bar{Q}\to \gamma\gamma)&=&-ic_{1}A_{\mu\nu}
\bar{Q}(\overrightarrow\partial_{\mu}-\overleftarrow\partial_{\mu})\gamma_{\nu}Q\\
\label{Leff}
\kern-0.5cm c_{1}&=& -e^{2}Q^{2}_{c,b}[(k_{1}-k_{2})^{2}/4 -m_{Q}^{2}]^{-2}\nonumber
\eea
With   the matrix element of
$\theta_{Q\mu\nu}$ between the vacuum and $\chi_{c0,2}$ or $\chi_{b0,2}$
given by ($Q^{2}=M^{2}$):
\bea
<0|\theta_{Q\mu\nu}|\chi_{0}> &=& T_{0}M^{2}(-g_{\mu\nu} + Q_{\mu}Q_{\nu}/M^{2}),\nonumber\\
<0|\theta_{Q\mu\nu}|\chi_{2}> &=& -T_{2}M^{2}\epsilon_{\mu\nu}.
\label{Tmn}
\eea
where $\epsilon_{\mu\nu} $ is the polarization tensor for  $\chi_{2}$
state, we obtain the two-photon decay amplitude in a  simple manner:
\bea
{\cal M}(\chi_{0}\to \gamma\gamma)&=& -e^{2}Q^{2}_{c,b}\frac{T_{0}A_{0}}{[M^{2}/4 +m_{Q}^{2}]^{2}}\\
\label{chi0}
{\cal M}(\chi_{2}\to \gamma\gamma)&=& -e^{2}Q^{2}_{c,b}\frac{T_{2}A_{2}}{[M^{2}/4 +m_{Q}^{2}]^{2}}
\label{chi2}
\eea
where
\bea
A_{0} &=&({3\over 2})M^{2}(M^{2}\epsilon_{1}\cdot\epsilon_{2}-2\epsilon_{1}\cdot k_{2}\epsilon_{2}\cdot k_{1})\\
\label{A0}
A_{2}&=&M^{2}\epsilon_{\mu\nu}[M^{2}\epsilon_{1\mu}\epsilon_{2\nu}
-2(\epsilon_{1}\cdot k_{2}\epsilon_{2\mu}k_{1\nu} + \epsilon_{2}\cdot
k_{1}\epsilon_{1\mu}k_{2\nu}\nonumber \\
&& + \epsilon_{1}\cdot \epsilon_{2}k_{1\mu}k_{2\nu})]
\label{A2}
\eea
The above expressions agree with the well-known non-relativistic
calculation of \cite{Barbieri}. The HQSS relation  $T_{2}=\sqrt{3}T_{0}$  
is obtained by \cite{Kuhn} in a calculation of the two-photon 
decays of $P$-wave quarkonium $\chi_{J}, J=0,2$ states
using the Bethe-Salpeter wave function and the relativistic spin
projection operators given in this reference and in \cite{Guberina}  
which is a precursor 
of the recent HQSS formulation of radiative decays of heavy 
quarkonium\cite{Lansberg,Lansberg2,Pham}. This method allows one to compute the
matrix element of derivative operator, like the energy-momentum tensor
in a bound state description of $P$-wave heavy quarkonium state, from
which HQSS relations could be obtained. However, for QCD sum rules calculation
or lattice simulation, non-derivative operator is simpler to
compute. Thus instead of working with the energy-momentum $\theta_{Q\mu\nu}$
operator, one could work with the trace $\theta_{Q\mu\mu}$ which, by
applying Dirac equation, becomes a  $\bar{Q}Q$ scalar density: 
\be
\theta_{Q\mu\mu}= 2m_{Q}\bar{Q}Q
\label{QQ}
\ee
and
\be
\bar{v}(p_{2})T_{\mu\mu}u(p_{1}) = 2m_{Q}\bar{v}(p_{2})u(p_{1})
\label{Tmm}
\ee
Then the problem of computing the two-photon or two-gluon decays of
$P$-wave  quarkonium $\chi_{c0,2}$ and $\chi_{b0,2}$ states is reduced to
computing the decays constants $f_{\chi_{c0}}$ or $f_{\chi_{b0}}$ 
states, defined as ($\chi_{0,2}$ denote here both $\chi_{c0,2}$ and $\chi_{b0,2}$ states):
\be
<0|\bar{Q}Q|\chi_{0}> = m_{\chi_{0}}f_{\chi_{0}}
\label{fchi}
\ee
Comparing Eq.(\ref{Tmn}) with Eq.(\ref{fchi}), we find:
\be
 T_{0}= {f_{\chi_{0}}\over 3}
\label{C0}
\ee
where we have neglected the binding energy\cite{Kuhn} $b=2m_{Q}- M$ and putting
$m_{Q}=M/2$. This agrees with the bound state calculations 
of \cite{Kuhn} and a
direct calculation of $\theta_{Q\mu\nu}$ and  $<0|\bar{Q}Q|\chi_{0}>$ 
using the expressions  Eq.(24-26)  of \cite{Guberina}. The point we
would like to stress here is that the local operator expansion allows us to
compute the two-photon and two-gluon decay amplitudes of $\chi_{c0,2}$
directly in terms
of the $f_{\chi_{c0}}$ decay constant, without using the wave function and
its derivative at the origin, as with that of  $\eta_{c}$ given in
terms of $f_{\eta_{c}}$\cite{Lansberg}.

 Another quantity of physical interest is the decay constant $f_{\chi_{1}}$
 of the $P$-wave $1^{++}$  $\chi_{c1}$ state which enters, for example,
in  $B\to \chi_{c1} K$\cite{Song,Beneke} and 
$B\to \chi_{c1} \pi$\cite{Wang} decays. Using expressions
Eq.(24-26) in \cite{Guberina} for $\chi_{0,1}$ states, we find in terms of
the derivative of the $P$-wave spatial wave function at the origin 
${\cal R}_{1}^{\prime}(0) $ :
\bea
 f_{\chi_{0}}&=& 12\sqrt{{3\over (8\pi m_{Q})}}\left({{\cal R}_{1}^{\prime}(0)\over M}\right)\nonumber\\
 f_{\chi_{1}} &=& 8\sqrt{{9\over (8\pi m_{Q})}}\left({{\cal R}_{1}^{\prime}(0)\over M}\right)
\label{f02}
\eea
which gives $f_{\chi_{1}}={\sqrt{3}\over 2}f_{\chi_{0}} $. 
Comparing with the $S$-wave singlet pseudo-scalar quarkonium decay constant
$f_{P}$\cite{Pham} ($M\simeq 2m_{Q}$):
\be
f_{\eta_{c}} = \sqrt{\frac{3}{32\,\pi\,m_{Q}^{3}}}\,{\cal R}_{0}(0)\, (4\,m_{Q}) \ , \qquad
\label{fP}
\ee
we have~:
\be
f_{\chi_{c0}} = 12\left({{\cal R}^{\prime}_{1}(0)\over {\cal R}_{0}(0)M}\right)f_{\eta_{c}}
\label{fSP}
\ee
where ${\cal R}_{0}(0) $ is the $S$-wave spatial wave function at the origin.

The two-photon decay rates of $\chi_{c0,2}$, $\chi_{b0,2}$ states can now
be obtained in terms of the decay constant $f_{\chi_{0}}$ . We find, 
either by using  Eq.(\ref{f02}) or  directly Eq.(\ref{C0}) for $T_{0}$:
\bea
&&\kern-0.9cm \Gamma_{\gamma\gamma}(\chi_{c0})=\kern-0.2cm \frac{4 \pi Q_c^4
   \alpha^2_{em}M_{\chi_{c0}}^{3} f_{\chi_{c0}}^2}{(M_{\chi_{c0}}+ b)^{4}}\left[1 + B_{0}(\alpha_{s}/\pi)\right],
\label{R0}\\
&&\kern-0.9cm \Gamma_{\gamma\gamma}(\chi_{c2})=\kern-0.2cm \left({4 \over 15}\!\right)\frac{4 \pi Q_c^4 \alpha^2_{em}M_{\chi_{c2}}^{3} f_{\chi_{c0}}^2}{(M_{\chi_{c2}}+ b)^{4}}\left[1 + B_{2}(\alpha_{s}/\pi)\right]
\label{R2}
\eea
where $B_{0}= \pi^{2}/3 -28/9$ and $B_{2}=  -16/3$ are NLO
QCD radiative corrections\cite{Barbieri2,Kwong,Mangano}. 
It is interesting to note that, the expression for the $\chi_{c0}$ two-photon
decay rate  is similar to that for $\eta_{c}$\cite{Lansberg}:
\be
 \Gamma_{\gamma\gamma}(\eta_{c})= \frac{4 \pi Q_c^4
   \alpha^2_{em}M_{\eta_{c}} f_{\eta_{c}}^2}{(M_{\eta_{c}}+
   b)^{2}}\left[1 -{\alpha_{s}\over \pi}{(20-\pi^{2})\over 3}\right]
\label{Retac}
\ee
In the same manner, we have~, for the two-gluon decays~:
\bea
&&\kern-0.9cm \Gamma_{gg}(\chi_{c0})= \left({2\over 9}\right)\frac{4 \pi 
   \alpha^2_{s}M_{\chi_{c0}}^{3} f_{\chi_{c0}}^2}{(M_{\chi_{c0}}+ b)^{4}}[1 + C_{0}(\alpha_{s}/\pi)],
\label{R0gg}\\
&&\kern-0.9cm \Gamma_{gg}(\chi_{c2})= \left({4 \over 15}\right)\!\left({2\over 9}\right)\!\frac{4 \pi  \alpha^2_{s}M_{\chi_{c2}}^{3} f_{\chi_{0}}^2}{(M_{\chi_{c2}}+ b)^{4}}[1 + C_{2}(\alpha_{s}/\pi)]
\label{R2gg}
\eea
where $C_{0}= 8.77$ and $C_{2}= -4.827$ are NLO 
QCD radiative corrections\cite{Barbieri2,Kwong,Mangano}. For comparison,
the expression for $\Gamma_{gg}(\eta_{c}) $ is similar:
\be
 \Gamma_{gg}(\eta_{c})=\left({2\over 9}\right) \frac{4 \pi
   \alpha^2_{s}M_{\eta_{c}} f_{\eta_{c}}^2}{(M_{\eta_{c}}+
   b)^{2}}\left[1 +4.8{\alpha_{s}\over \pi}\right]
\label{Retacgg}
\ee

We have seen that,  the usual expression for the
decay rate $\Gamma_{\gamma\gamma}(\chi_{0})$ in terms of 
${\cal  R}_{1}^{\prime}(0)$ is now reduced to the simple form Eq.(\ref{R0})
by using Eq.(\ref{f02}). We note also that   Eq.(\ref{fSP}) shows that
$f_{\chi_{c0}} $ becomes comparable to $f_{\eta_{c}} $,  even though, 
in general, $T_{0}$ and $f_{\chi_{c0}} $
are of the order $O(q/M)$ compared with $f_{\eta_{c}} $. 

\begin{table}[h]
\begin{tabular}{|c|c|c|c|}
\hline
 Reference
 &$\Gamma_{\gamma\gamma}(\chi_{c0})$(keV)&$\Gamma_{\gamma\gamma}(\chi_{c2})$(keV)&$R={\Gamma_{\gamma\gamma}(\chi_{c2})\over
 \Gamma_{\gamma\gamma}(\chi_{c0})}$ \\ 
\hline
Barbieri\cite{Barbieri}&$3.5$ &$0.93$ &$ 0.27$\\
Godfrey\cite{Godfrey}&$1.29$  &$0.46$ & $0.36$\\
Barnes\cite{Barnes}&$1.56$ &$0.56$&$0.36$ \\
Bodwin\cite{Bodwin} &$6.70\pm 2.80$&$0.82\pm 0.23$ &$0.12^{+0.15}_{-0.06}$ \\
Gupta\cite{Gupta} &$6.38$&$0.57 $ &$0.09 $\\
M\"unz\cite{Munz} &$1.39\pm 0.16 $ &$0.44\pm 0.14 $&$0.32^{+0.16}_{-0.12}$\\
Huang\cite{Huang} &$3.72\pm 1.10 $ &$0.49\pm 0.16 $&$0.13^{+0.11}_{-0.06}$\\
Ebert\cite{Ebert} &$2.90 $ &$0.50 $&$0.17$\\
Schuler\cite{Schuler} &$2.50 $ &$0.28 $&$0.11$\\
Crater\cite{Crater} &$3.34-3.96 $ &$0.43-0.74 $&$0.13-0.19$\\
Wang\cite{LWang} &$3.78 $ &$- $&$-$\\
Laverty\cite{Laverty} &$1.99-2.10 $ &$0.30-0.73 $&$0.14-0.37$\\
This work  &$5.00 $ &$0.70 $&$0.14$\\
\hline
\end{tabular}
\caption{ Potential model predictions for $\chi_{c0,2}$
 two-photon widths compared with this work.}
\end{table}
 In  the limit of $b=0$, the expressions 
for the two decay rates  are exactly the same, apart from the decay 
constant $f_{\chi_{0}}$ and $f_{\eta_{c}}$ and QCD radiative 
correction terms. The decay rate for $\chi_{c2}$ differs from that of
 $\chi_{c0}$ only by a HQSS factor.  Thus by comparing the expression for 
$\chi_{c0}$ and $\eta_{c}$ we could already have some estimate for the
$\chi_{c0}$ two-photon and two-gluon decay rates. For $f_{\chi_{c0}}$
of $O(f_{\eta_{c}})$, one would expect
 $\Gamma_{\gamma\gamma}(\chi_{c0})$ to be in the range of a few keV and that
$\Gamma_{gg}(\chi_{c0})$ is roughly of 
 the same size as  $\Gamma_{gg}(\eta_{c})$ obtained with the
 QCD sum rules values\cite{Novikov}~ for the decay constants~:
 $f_{\eta_{c}}=374\rm\,MeV$ and $f_{\chi_{c0}}=359\rm\,MeV$ which gives,
 with QCD radiative corrections (NLO value):
 $\Gamma_{\gamma\gamma}(\eta_{c})= 4.33\rm\,keV$, 
$\Gamma_{\gamma\gamma}(\chi_{c0})=5.0\rm\,keV$ and 
$\Gamma_{\gamma\gamma}(\chi_{c2})=0.70\rm\,keV$ to be compared,
with the latest CLEO result of $(2.53\pm 0.37\pm 0.26)\rm\,keV$ 
and $(0.60\pm 0.06\pm 0.06)\rm\,keV$; and the averages of all current 
measurements~:
$(2.31\pm 0.10\pm 0.12\rm\,keV$, $0.51\pm 0.02\pm 0.02)\rm\,keV$,
respectively, for $\chi_{c0}$ and  $\chi_{c2}$ two-photon width\cite{CLEO}.
For $\eta_{c}$ the prediction from the sum rules value of $f_{\eta_{c}}$
mentioned above  is slightly less than 
the NLO value of $5.34\rm\,keV$ obtained with HQSS and is  
more or less in agreement with experiment. Similarly, the prediction 
for $\chi_{c0}$ from the sum rules value for $f_{\chi_{c0}}$
is however almost twice  the CLEO value, but 
possibly with large theoretical uncertainties in sum rules calculation 
for $f_{\chi_{c0}}$, as to be expected. For comparison, we note 
that the Cornell potential model gives 
$f_{\chi_{c0}}= 338\rm\,MeV$\cite{Quigg}. Also a recent QCD sum rules
calculation\cite{Colang} gives
$f_{\chi_{c0}}=510\pm 40\rm\,MeV$ which implies a still larger $\chi_{c0}$
two-photon decay rates. Various potential model calculations give 
$\Gamma_{\gamma\gamma}(\chi_{c0})$ in the range $1.2-6.7\  \rm\,keV$
and $\Gamma_{\gamma\gamma}(\chi_{c2})$ in the range $0.28-0.93\  \rm\,keV$
as shown in table I. From the above expressions for the decay rates, the
two-photon branching ratios for $\chi_{c0}$ and  and $\eta_{c}$
would be the same in the absence of QCD radiative corrections(the 
two-photon $\chi_{c2}$ branching ratios is smaller by $20\%$ 
with ${\cal  B}_{\gamma J/\psi(1S)}(\chi_{c2})= (20\pm 1.0)\%$\cite{PDG}). 
With QCD radiative correction, the predicted branching ratios are~, for 
$\alpha_{s}=0.28$: 
${\cal  B}_{\gamma\gamma}(\eta_{c})= 2.90\times 10^{-4}$, 
${\cal  B}_{\gamma\gamma}(\chi_{c0})= 3.45\times 10^{-4}$, and 
${\cal  B}_{\gamma\gamma}(\chi_{c2})= 2.55\times 10^{-4}$ which are very close
to the measured value of $(2.4^{+1.1}_{-0.9})\times 10^{-4} $, 
$(2.35 \pm 0.23)\times 10^{-4} $, and $(2.43 \pm 0.18)\times 10^{-4} $,
respectively\cite{PDG}. This shows that QCD radiative corrections 
are important in bringing the predictions close to experiments.

For the excited state $2P$ $\chi_{c0,2}$ states, there has been
observation of the $\chi^{\prime}_{c2}$ state above $D\bar{D}$
threshold, the $Z(3930)$ state, at $M=(3928\pm 5\pm 2)\rm\,MeV$ by the Belle 
Collaboration\cite{Belle} which gives 
$\Gamma_{\gamma\gamma}(\chi^{\prime}_{c2})\times {\cal
  B}(D\bar{D})=(0.18\pm 0.05\pm 0.03)\rm\,keV$ which implies 
$\Gamma_{\gamma\gamma}(\chi^{\prime}_{c2})\simeq
(0.18\pm 0.04)\rm\,keV$\cite{Colangelo2}. This would imply  
$\Gamma_{\gamma\gamma}(\chi^{\prime}_{c0})\simeq
(1.30\pm 0.3)\rm\,keV$, and $f_{\chi^{\prime}_{c0}}\simeq 195\rm\,MeV$, 
comparable with the HQSS value of $279\rm\,MeV$ for
$f_{\eta_{c}^{\prime}}$\cite{Lansberg}. One thus expects that 
$\Gamma_{gg}(\chi^{\prime}_{c0})$ in the range $5-10 \rm\,MeV$.

For $\chi_{b0,2}$ the same potential model
calculation quoted in table I  gives the $\chi_{b0,2}$ two-photon 
width  about $1/10$ of that for $\eta_{b}$ which implies   
$f_{\chi_{b0}}\approx (1/3)f_{\eta_{b}}$, smaller  than the value
obtained from ${\cal R}_{0}(0) $ and ${\cal R}^{\prime}_{1}(0) $ with
the Cornell potential\cite{Quigg} which gives $f_{\chi_{b0}}= 0.46f_{\eta_{b}}$.
\bigskip
\section{Conclusion}
By using  local operator expansion, we show that the two-photon and 
two-gluon decays of the $P$-wave heavy quarkonium $\chi_{c0}$ and $\chi_{b0}$
state  can be obtained from the heavy quark energy-momentum tensor and 
its trace, a  $\bar{Q}Q$ scalar density. The decay rates can then be 
expressed in terms of  $f_{\chi_{c0}}$ and $f_{\chi_{b0}}$ decay
constant and are similar to that of $\eta_{c}$. Existing sum rules
calculation for $f_{\chi_{c0}}$ however produces a $\chi_{c0}$
two-photon width about $5\rm\,keV$, somewhat bigger than the 
CLEO measurement, but possibly with large theoretical uncertainties. It remains
to be seen whether a better determination of $f_{\chi_{c0}}$
could bring the $\chi_{c0,2}$ two-photon decay rates closer to experiments
or higher order QCD radiative corrections and large relativistic corrections
are needed to explain the data.

\bigskip

\begin{center}
{\bf Acknowledgments} \end{center}

This work was supported in part by the EU 
contract No. MRTN-CT-2006-035482, "FLAVIAnet", by a Francqui fellowship 
of the Belgian American Educational Foundation and
by the U.S. Department of Energy under contract number DE-AC02-76SF00515.

\end{document}